\documentclass[aps,pre,reprint,amssymb]{revtex4-1}
\usepackage{graphicx}
\usepackage{bm}
\usepackage{amssymb}
\usepackage{color}
\usepackage{changes}

\newcommand{\eps}{\varepsilon}
\newcommand{\Wa}{W_{\mathrm{act}}}
\newcommand{\Wp}{W_{\mathrm{proj}}}
\newcommand{\Wo}{W_{\mathrm{o}}}

\begin{document}
\title{Birth and decay of tensional wrinkles in hyperelastic sheets}
\author{Andreea Panaitescu,$^1$ Meng Xin,$^2$ Benny Davidovitch,$^2$ Julien Chopin,$^3$ and Arshad Kudrolli$^1$}
\affiliation{$^{1}$Department of Physics, Clark University, Worcester, MA 01610, USA}
\affiliation{$^{2}$Department of Physics, UMass Amherst, Amherst, MA 01003, USA,}
\affiliation{$^3$Instituto de F\'isica,  Universidade Federal da Bahia, Salvador-BA 40170-115, Brazil}

\date{\today}

\begin{abstract}
We demonstrate with experiments that wrinkling in stretched latex sheets occur over finite strains, and that their amplitudes grow and then decay to zero over a greater range of applied strains compared with linear elastic materials. The wrinkles occur provided the sheet is sufficiently thin compared to its width, and only over a finite range of length-to-width ratios.  We show with simulations that the Mooney-Rivlin hyperelastic model describes the observed growth and decay of the wrinkles in our experiments. The decrease of wavelength with applied tension is found to be consistent with a far-from-threshold scenario proposed by Cerda and Mahadevan in 2003. However, the amplitude is observed to decrease with increasing tensile load, in contrast with the prediction of their original model. We address the crucial assumption of {\it collapse of compressive stress}, as opposed to collapse of compressive strain, underlying the far-from-threshold analysis, and test it by measuring
 the actual arc-length of the stretched sheet in the transverse direction and its difference from the width of a planar projection of the wrinkled shape.  
Our experiments and numerical simulations indicate a complete {\it collapse of the compressive stress}, and reveal that a proper implementation of the far-from-threshold analysis is consistent with the non-monotonic dependence of the amplitude on applied tensile load observed in experiments and simulations. Thus, our work support and extend far-from-threshold analysis to the stretching problem of rectangular hyperelastic sheets.  
\end{abstract}
\maketitle

\section{Introduction}

Wrinkling instabilities are well known to occur in thin elastic sheets due to a subtle competition between bending and stretching energetics under a variety of applied boundary conditions~\cite{Audoly2010,Davidovitch2011,Chopin2013}. A particularly interesting example is a thin rectangular elastic sheet which when clamped at its two short ends and stretched, displays periodic wrinkles oriented parallel to the direction of applied tensile strain $\varepsilon$~\cite{Friedl2000}. It has been proposed that hindered Poisson contraction near the clamped edges leads to the formation of wrinkles above a critical value of $\varepsilon = \varepsilon_c$~\cite{Cerda2002}, which eventually extend throughout the whole sheet when $\varepsilon \gg \varepsilon_c$. Subsequent scaling arguments inspired by the classical F\"oppl-von K\'arm\'an (FvK) equations for thin Hookean (i.e. linear elastic) sheets were used to predict the behavior of the observed wrinkle amplitude and wavelength with applied strain~\cite{Cerda2003}. However, this approach has been criticized as introducing an ad hoc geometrical constraint that a transverse strain is eliminated by the formation of wrinkles~\cite{Kim2012}, leading to incorrect conclusions on the growth of the wrinkle amplitude above threshold~\cite{Healey2013}. In fact, the prediction of an unbounded growth of the wrinkles amplitude has been found to be inconsistent with results of simulations with neo-Hookean models~\cite{Zheng2009,Nayyar2011,Kim2012a,Healey2013,Taylor2014}. In these simulations, the wrinkles are found to grow, decay, and finally disappear with increasing strain at large but finite values. 
Interestingly, the same scaling analysis was found to yield consistent results for the evolution of the wrinkle wavelength with strain~\cite{Cerda2002,Nayyar2011,Kim2012}.

More recent simulations~\cite{Li2016,Fu2018} have noted that the neo-Hookean and Mooney-Rivlin models of elastic materials predict wrinkles at higher strains than first reported with the Saint Venant-Kirchhoff model~\cite{Healey2013}, in addition to qualitative differences in the region of parameter space where wrinkles are expected to appear. The situation with respect to experiments and reported observations remains even less clear. Some studies with silicone rubber sheets appear to show wrinkles over a large range of applied strain where simulations find wrinkles to flatten out completely~\cite{Zheng2009}. Experimental errors, material imperfections and orthotropicity have been proposed as explanations for this apparent discrepancy~\cite{Taylor2014,Sipos2016}. In studies with polyethylene sheets, it was found that the observed wrinkles also persist to larger strains compared with finite element simulations implementing hyperelastic constitutive models~\cite{Nayyar2011}. Furthermore, it was shown that these materials are strain rate-dependent and deform plastically as strain is increased~\cite{Nayyar2011}.  

The far-from-threshold analysis proposed in Ref.~\cite{Cerda2003} has been pursued and developed further in studies that addressed simpler systems, yielding analytic progress and elucidating key aspects of this physical approach \cite{Davidovitch2011,Chopin2015,Paulsen2016,Vella2018b,Davidovitch2019}. Specifically, it was noted that rather than elimination of compressive strain that was invoked in Ref.~\cite{Cerda2002}, wrinkling of thin sheets in the far from threshold regime completely relaxes the in-plane compressive stress in the solid. For the stretching problem addressed here, this implies that the ratio between the wrinkle amplitude and wavelength is determined by the difference between the measured arc-length of the wrinkled width of the sheet and the planar projection of this shape, in a manner that is different from the one obtained in the original version of the far-from-threshold model~\cite{Cerda2003}. However, to our knowledge, an experimental assessment of the relevant assumption underlying a proper implementation of far-from-threshold analysis -- collapse of compressive strain {\emph{versus}} stress -- has not been conducted. 

Indeed, experimental investigations remain few, and a quantitative agreement between experiments and numerical predictions in case of elastic materials has not been demonstrated in spite of recent efforts to implement more elaborate numerical models that include geometrical nonlinearities, and various nonlinear rate-dependent and rate-independent constitutive laws for stretching and bending deformations~\cite{Nayyar2014,Li2016,Fu2018,Xin2018}. Thus, a combined numerical and experimental investigation of tension-induced wrinkles on simpler materials is highly needed to clarify the nature of the wrinkles and their persistence to higher strains in the case of elastic sheets. 

Here, we investigate tensional wrinkling patterns by using a highly stretchable latex sheets which are shown to be hyperelastic under the application of a large range of strain. In this case, we show that wrinkles indeed arise and flatten out at finite strains, in apparent contrast with the proposed amplitude growth~\cite{Cerda2003}, but which are also nonetheless quantitatively different from those reported by recent numerical work on nonlinear plate models using Hookean~\cite{Healey2013,Taylor2014} and hyperelastic materials~\cite{Nayyar2011,Li2016,Fu2018,Wang2019}. We observe that the wrinkles in fact persist to far greater strains, and show that the observed evolution is consistent with simulations using the Mooney-Rivlin model after the material parameters are matched appropriately. Thus, we show that the nature of the material nonlinearity affects significantly the region in the parameter space where wrinkling occurs. We then show that the growth and decay of the measured amplitude is inconsistent with 
the assumption of collapse of transverse strain~\cite{Cerda2002}. Rather, we show it to be consistent with a far-from-threshold approach derived from assuming a collapse of compressive stress. Our experimental and numerical work not only validates this assumption, but also shows its applicability to hyperelastic materials in capturing the growth of the wrinkle amplitude along with its wavelength. 
 
\section{Experimental System} 

\begin{figure}
\begin{center}
\includegraphics[width = 6.5cm]{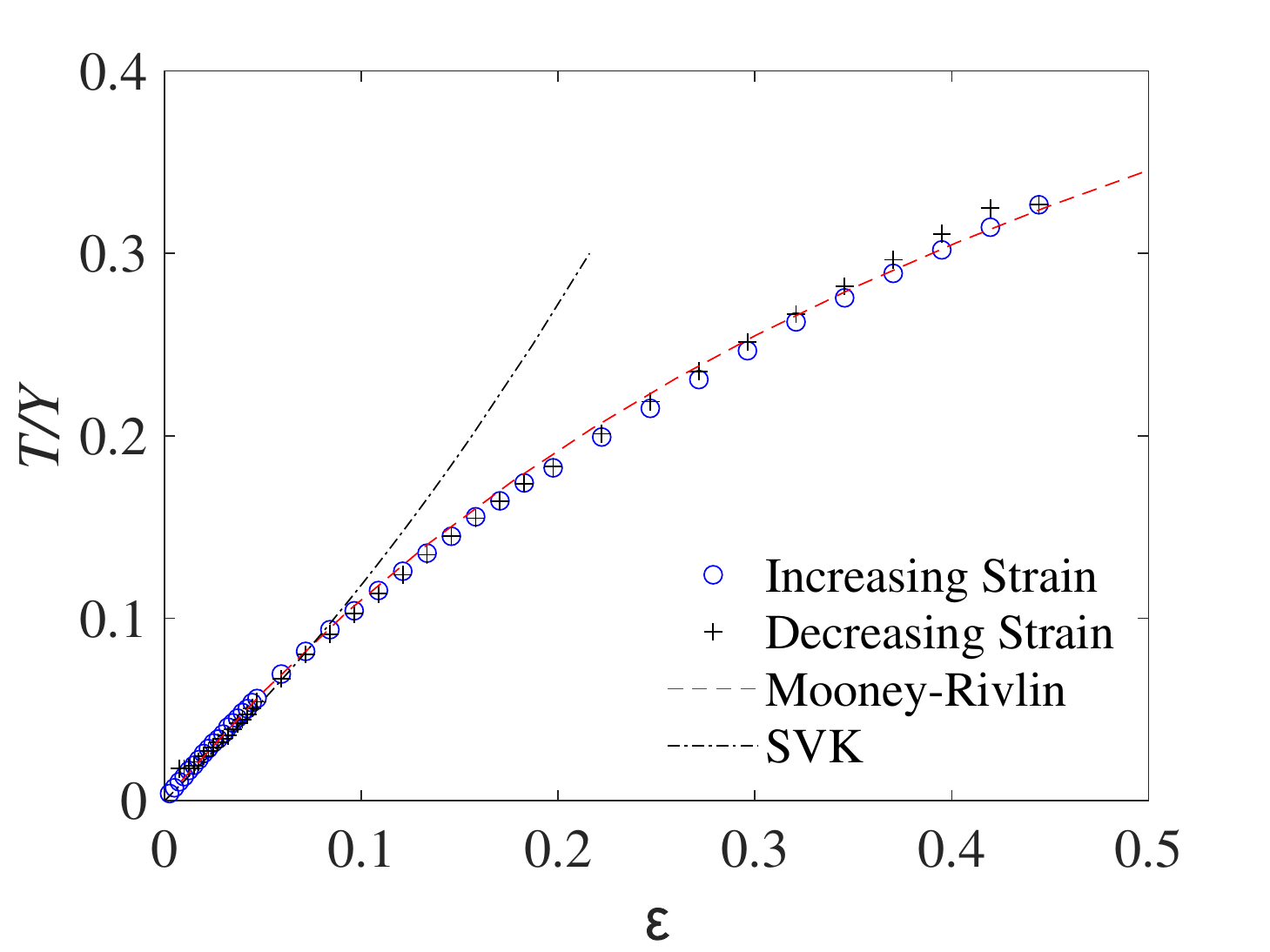}
\end{center}
\caption{Applied stress $T$ normalized by the Young's modulus $Y$ versus measured strain for latex while strain is increased and decreased. The differences are within the experimental error which is of the order of the size of the symbols.  The data is well described by the Mooney-Rivlin hyperelastic model for soft elastic solids. The measured $Y = 1$\,MPa. 
}   
\label{fig:props}
\end{figure}
The experimental system consists of a latex sheet with initial length $L_o$, width $\Wo$, and thickness $t_o$, mounted horizontally between two rigid parallel clamps which are further attached to translating stages, similar to the one used previously to study the shape and strain experienced by unwrinkled thick elastic sheets~\cite{Chopin2018}. The clamps are moved axially through prescribed separation distances $L$, while maintaining their orientation, resulting in an applied longitudinal strain $\eps = \Delta L/L_o = (L -L_o)/L_o$. Experiments were performed with latex sheets with thicknesses in the 140 to 500 micron range. The sheets are  steamed and ironed to remove any creases, and then used over a period of days. In order to obtain consistent results, care has to be taken in preparation of the sheets and in their handling, including storing the sheets in air-tight containers, until they are ready to use. If these precautions are not taken, rubber sheets 
become somewhat thinner and brittle over longer periods.  

To characterize the elastic properties of the sheets, complementary measurements were performed by stretching and relaxing sheets of the materials by varying the applied strain $\eps$. The measured nominal stress $T$ scaled by the Young's modulus $Y$, corresponding to the initial slope of the curve, is plotted as a function of $\eps$ in Fig.~\ref{fig:props} over a range of applied strains which is used to investigate wrinkling. The response does not show any significant plasticity, as can be seen from the close overlap of the points corresponding to when the strain is increased and decreased in Fig.~\ref{fig:props}. It can be also observed that the stress-strain curve is nonlinear above $\eps \approx 0.1$ compared to the linear elastic constitutive law used in the Saint Venant-Kirchhoff (SVK) model. Limited experiments were also performed with PDMS sheets, where similar results were observed, but are not reported here for simplicity of presentation. 

\begin{figure}
\begin{center}
\includegraphics[width = 8cm]{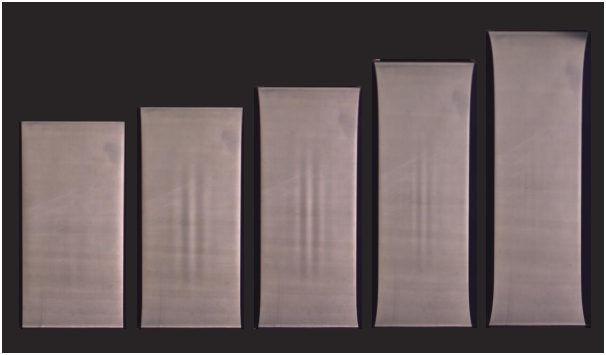}
\end{center}
\caption{Wrinkles are formed in the central region of a sheet ($L_o/\Wo = 2$ and $t_o/\Wo = 1 \times 10^{-3}$) which is clamped at two ends and then stretched  uniaxially. The applied strain $\eps = 0$, 0.1, 0.2, 0.32, and 0.46.}   
\label{fig:ima}
\end{figure}

The stress response of rubber is typically described with the Mooney-Rivlin (MR) model~\cite{treloar1975physics}. We implement the extension of the sheet with the same dimensions $L_o = 28$\,cm, $\Wo = 14$\,cm, and $t_o = 140$\,$\mu$m used in our experiments using the Mooney-Rivlin model in Abaqus. The fitting parameters in the MR model which describes the data are found to be $D_1 = 0$\,Pa corresponding to the volumetric response, $C_{01} = 1.58 \times 10^5$\,Pa, and $C_{10} = 6.21 \times 10^4$\,Pa, the material constants related to the distortional response. The corresponding curve is plotted in Fig.~\ref{fig:props}, and is observed to be in excellent agreement with the data. We also compare the curve with the SVK model, where we assume Young's modulus similar to the initial measured response of latex used, 
and an ideal incompressible material. However, systematic deviations are observed, confirming that the relevant hyperelastic response over applied strains of interest are indeed given by the MR model. Having established the appropriate hyperelastic model in the sheets used, we next describe the experiments and measurements on wrinkling of the sheet, before comparing with the corresponding simulations.  

\section{Wrinkling observations}

\begin{figure}
\begin{center}
\includegraphics[width=8cm]{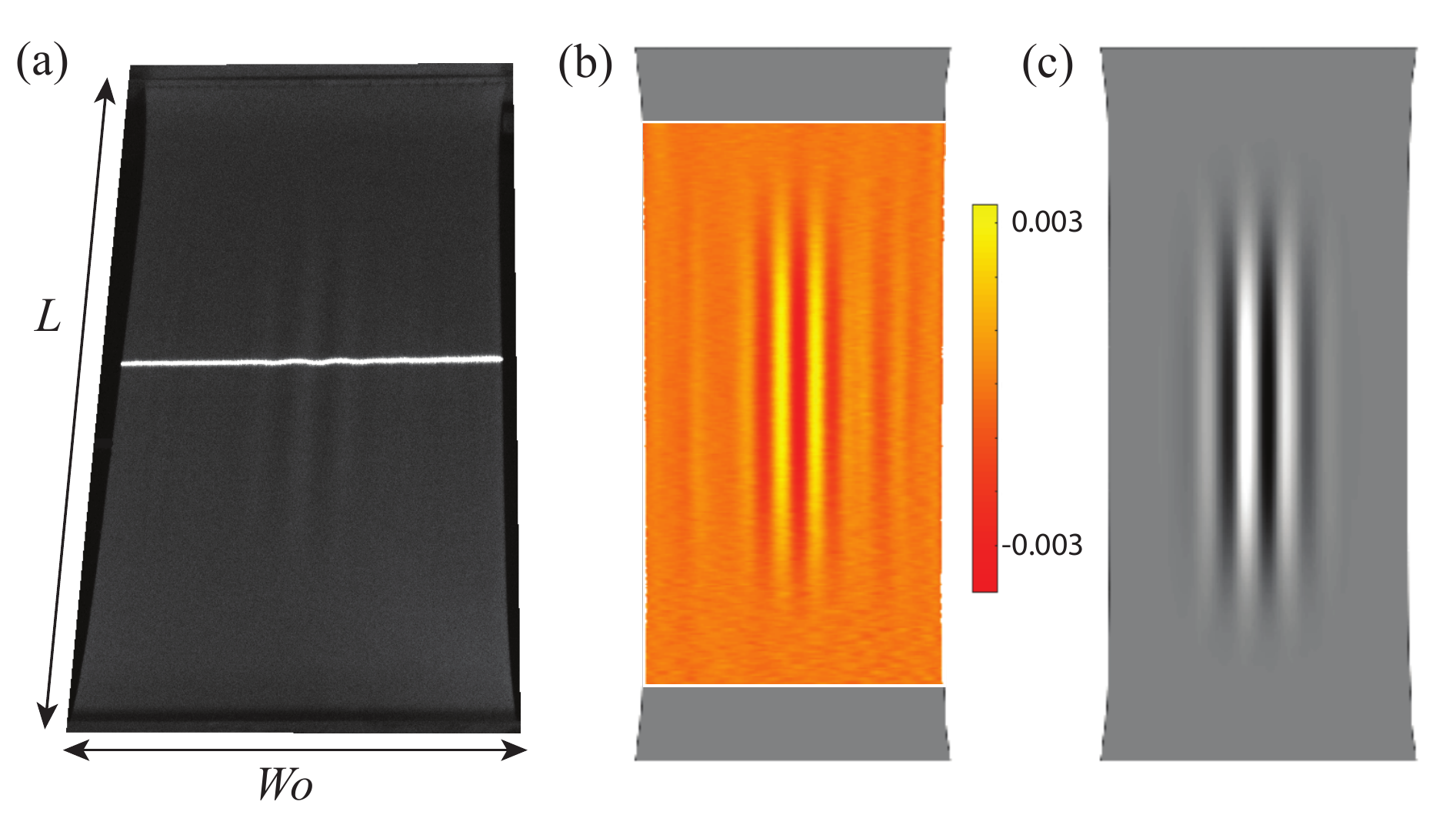}
\end{center}
\caption{(a) An image of the wrinkled sheet with a crosssection illuminated by a laser sheet. (b) A surface rendering of the scanned sheet with deflection mapped to the color map. The scanned area corresponds to 80\% of the central section of the sheet. (c) The corresponding simulated profile under the same conditions ($L_o/\Wo = 2$; $t_o/\Wo = 1 \times 10^{-3}$, $\eps = (L - L_o)/\Wo = 0.12$). }   
\label{fig:eg}
\end{figure}
Figure~\ref{fig:ima} shows a sequence of images of a planar rectangular latex sheet with $L_o = 28.0$\,cm, width $\Wo = 14.0$\,cm, and thickness $t_o = 140 \pm 1\,\mu$m as the applied longitudinal strain is increased. We choose the coordinate system ($x,y$) such that $x$ is long axis (along which tension is applied), and the origin ($0,0$) is the center of the rectangular sheet.  
Here, the sheet is viewed directly from above while being illuminated with a pair of lights to the side, and thus the regions with greater slope appear darker. One observes that wrinkles form at the central regions of the sheet and grow in amplitude before flattening out. Thus, we establish that wrinkles appear as an instability of a planar deformation, after a finite strain is applied, 
and then indeed disappear when sufficiently large strain is applied. Hence, our experimental results are qualitatively consistent with recent theoretical and simulation work which find that wrinkles only occur over a finite range of applied strains~\cite{Nayyar2011,Healey2013}. However, one can also observe that wrinkles persist to far greater strains than the range 0.1 to 0.23 reported previously with elastic sheets of similar dimensions with a wide range of models~\cite{Nayyar2011,Healey2013,Li2016}. 

\begin{figure}
\begin{center}
\includegraphics[width = 6.5cm]{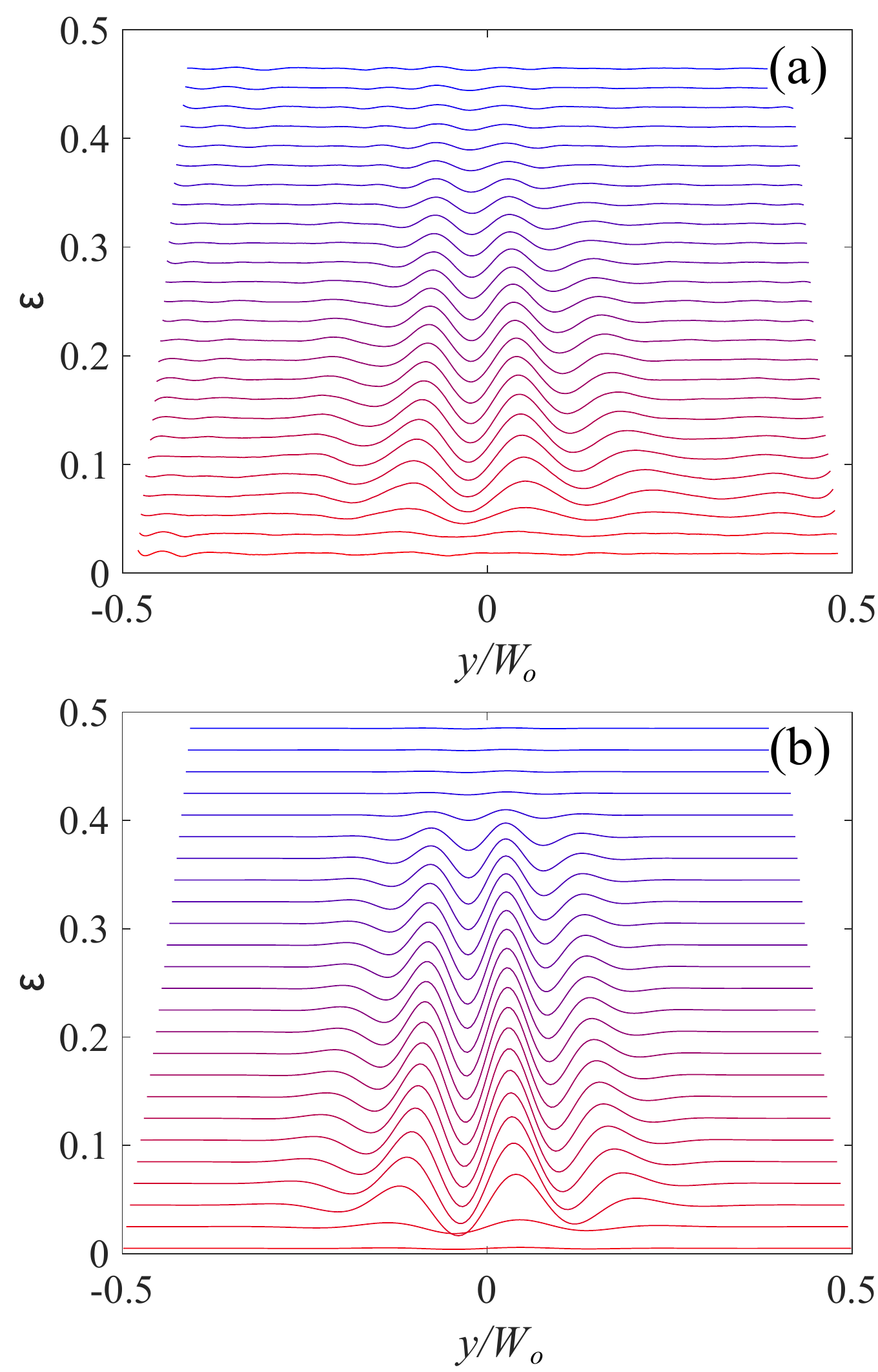}
\end{center}
\caption{(a) Plots of the measured amplitude profile of the sheet mid-way between the clamps in the experiments. The profiles are shifted up by a factor proportional the applied strain $\eps$ noted on the vertical axis. (b) The corresponding Mooney-Rivlin model profiles obtained with numerical simulations. The wrinkles appear to grow and decay similarly with increasing strain ($L_o/\Wo = 2$; $t_o/\Wo = 1 \times 10^{-3}$). }   
\label{fig:profiles}
\end{figure}

\subsection{Laser scanned wrinkle profiles}
To observe the growth and decay of the wrinkles with strain more quantitatively, we use laser profilometry~\cite{Blair2005}. In this technique, a thin vertical light sheet is used to illuminate the elastic sheet perpendicular to its surface, and the resulting illuminated intersection is imaged from an angle. A sample image is shown in Fig.~\ref{fig:eg}(a). The deflection of the illuminated sheet from a straight line is obtained using standard image processing, and after appropriate calibration, the amplitude of deflection of the sheet is detected to within a few microns, well within the thickness of the sheet. A rendering of the scan of the wrinkled sheet, performed at $0.01L_o$ intervals along the length of the sheet, is shown in Fig.~\ref{fig:eg}(b). The deflection amplitude is mapped to the color bar, which is between $\pm 3.5 \times 10^{-3} \Wo$, or about $\pm 2.5t_o$. Here, the central 80\% of the sheet is scanned because the flat areas near the bounded edges are not obtained as accurately due to eclipsing by the clamping armature. 

We plot the deflection amplitude of the sheet across the width of the sheet midway between the clamped edges in Fig.~\ref{fig:profiles}(a) to further illustrate the development of the wrinkles with quantitative measurements. Here, we have plotted each profile shifted up by a factor proportional to the applied strain $\eps$ as noted on the left axis. As $\eps$ is increased, we observe that the wrinkles increase rapidly in amplitude, before decreasing more slowly to essentially zero at the greatest strains applied, confirming the trends shown in the images in Fig.~\ref{fig:ima}. Thus, we clearly demonstrate that a cleanly prepared planar sheet wrinkles at finite applied strain, and then flatten out as applied strain is increased further.

\subsection{Numerical Simulations with Mooney-Rivlin Model} 
We now discuss the results of the simulations performed with the MR model with parameters matched to capture the stress-strain response discussed in the previous section.  Figure~\ref{fig:eg}(c) shows an example of the simulated sheet surface corresponding to the example shown in Fig.~\ref{fig:eg}(b). We observe that wrinkles obtained in the simulations appear similar and are localized in the center of the sheet as in the experiments. 

To compare the overall development over the range of applied strains with the experiments, we also plot the profiles across the midpoint between the clamps in the simulations in Fig.~\ref{fig:profiles}(b). There it can be noted that the profiles are antisymmetric in both the experiments as well as the simulations. We also observe that the overall development and relaxation of the wrinkles 
appear to be similar as a function of strain.  We next make quantitative comparisons between the wrinkle amplitude and wavelength observed in experiments and simulations.

\subsection{Amplitude Growth Comparisons} 
\begin{figure}
\begin{center}
\includegraphics[width = 6.5cm]{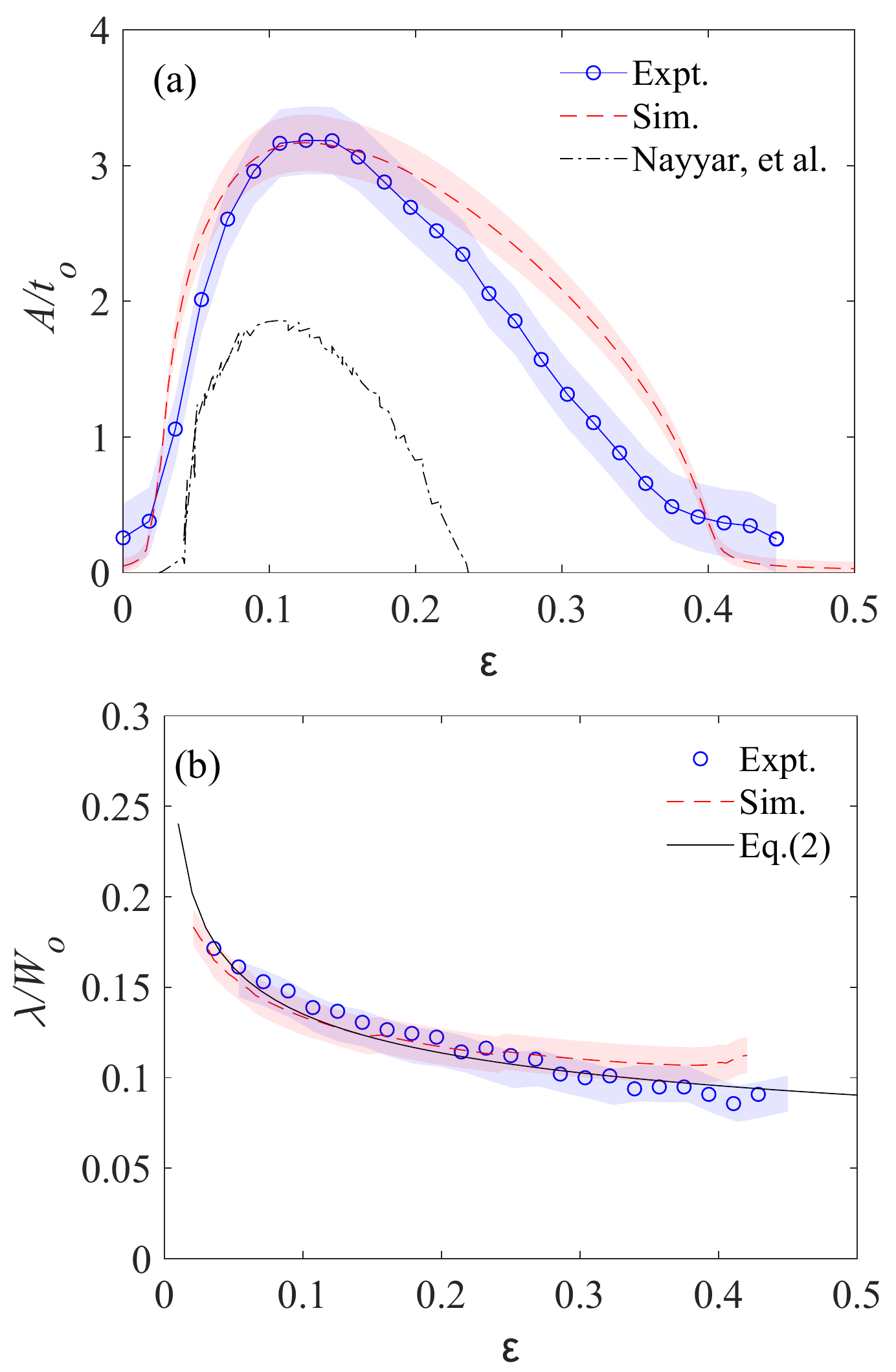}
\end{center}
\caption{
(a) The measured amplitude versus applied strain compared with amplitudes determined using simulations of the MR model ($L_o/\Wo = 2$ and $t_o/\Wo = 10^{-3}$.) The shaded area corresponds to the experimental error.  The experimental data is in agreement with MR simulations. The neo-Hookean model simulations by Nayyar, {\it et al.}~\cite{Nayyar2011} (line) are also shown for reference. (b) Measured wavelengths with $L_o/\Wo = 2$ and $t_o/\Wo = 10^{-3}$  as a function strain compared with Eq.~\ref{eq:CM-wave} and simulations with the MR model.
}
\label{fig:strain}
\end{figure}
We obtain the maximum wrinkle amplitude $A$ in each profile shown in Fig.~\ref{fig:profiles}(a) in case of the experiments, and Fig.~\ref{fig:profiles}(b)  in case of the simulations. The measured values of $A$ are normalized by the thickness of the sheet $t_o$ and plotted as a function of applied strain in Fig.~\ref{fig:strain}(a). The errors due to the material variation in the sheet thickness and the calibration of the profiles are shown by the shaded areas.  The experiments and our simulations with the MR model are in good overall agreement, not only 
for the peak amplitude and the strain at which it is reached, but also for the range over which the wrinkle amplitudes decay to zero. This agreement, along with overall trends in the surface profiles shown in Fig.~\ref{fig:eg} and Fig.~\ref{fig:profiles} show that the MR model is an excellent descriptor of the elastic response of latex sheets, and specifically the wrinkle patterns in them.

We also compare the observed amplitude against the simulations reported by Nayyar, {\it et al.}~\cite{Nayyar2011} using the neo-Hookean (NH) hyperelastic model for perspective. We observe that while the experimental data is overall consistent with the increase and then decrease observed in these simulations, the measured amplitude extends over a significantly higher strain~$\eps$ besides being systematically greater. 
Li and Healey~\cite{Li2016} and Fu, {\it et al.}~\cite{Fu2018} have also reported simulation results using the Mooney-Rivlin model with a sheet with similar dimensions where the maximal strain at which wrinkles disappear was around 0.3, notably smaller than the experimentally observed value, and our own simulations.  This can occur because the material constants used in those simulations may be different, even if the sheet dimensions used are the same.

Thus, it appears that the persistence of wrinkles to higher strains is strongly affected by material nonlinearities~\cite{Fu2018}. Interestingly, the SVK plate model which implements a Hookean thin plate model using a geometric exact stretching energy and FvK-like bending energy is also able to capture qualitatively the disappearance of the wrinkles at large strain~\cite{Healey2013}, but at much lower strains. Thus, nonlinearities stemming from the strain-displacement relation have to be included to quantitatively model the tensional wrinkles in hyperelastic sheets. 

\subsection{Aspect Ratio dependence}
\begin{figure}
\begin{center}
\includegraphics[width = 6.5cm]{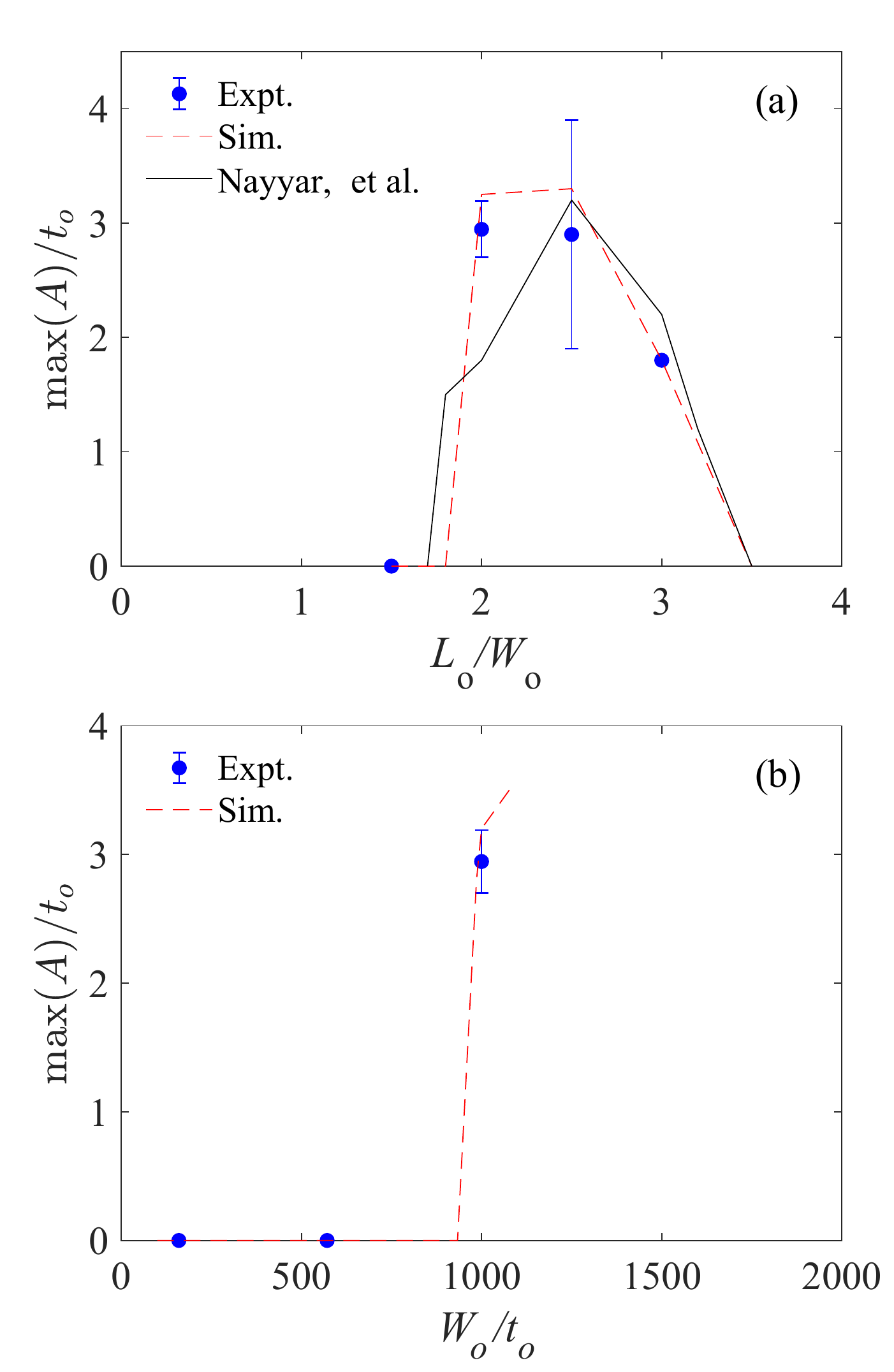}
\end{center}
\caption{The maximum amplitude versus strain for various $L_o/\Wo$ (a) and $W_o/t_o$ (b) when $t_o/\Wo = 1 \times 10^{-3}$.} 
\label{fig:max}
\end{figure}

According to Li and Healey's analysis of the FvK equations~\cite{Li2016}, the wrinkling instability can occur even below $L_o/\Wo = 1.5$ for $t_o/\Wo = 10^{-3}$ at sufficiently large $\eps$ as well as for arbitrarily large $L_o/\Wo$. However, their analysis 
of the NH, SVK and MR models find only an ``island" in the parameters space where wrinkling occurs over the range $1.5 \lesssim L_o/\Wo \lesssim 2.6$ in case of the NH and MR models, and slightly higher values for SVK model.  

In order to elucidate this issue, we performed further experiments and simulations with ratios $L_o/\Wo = 1.5, 1.8, 2, 2.5, 3.0$, and 3.5, while holding $\Wo$ and $t_o$ at constant values. The maximum observed amplitude max($A$) normalized by the sheet thickness is plotted in Fig.~\ref{fig:max}(a). No wrinkles are observed in case of $L_o/\Wo = 1.5$ in our experiments and simulations with MR model. 

Further, we also compare the measured occurrence of wrinkling with sheets with various thickness-to-width ratio ($t_o/\Wo$), for $L_o/\Wo = 2$. We observe that the wrinkles occur over a similar range of parameters in our experiments and MR simulations, and the wrinkle amplitudes are also consistent within experimental errors.  

Comparing our results with previous reports, we note that wrinkling is observed in the case of $L_o/\Wo = 3$, at which value no wrinkles are predicted according to the analysis of Li and Healey~\cite{Li2016}.  In comparing with the NH simulations by Nayyar, {\it et al.}~\cite{Nayyar2011}, a slightly higher cutoff is observed in $L_o/\Wo$, below which wrinkles are not observed. But, a similar upper cutoff is observed in $L_o/\Wo$ above which wrinkles are not observed. Thus, we find that while the qualitative trends in terms of an increase and decrease of wrinkling amplitude occurs in our experiments and various elastic models used in the simulations reported previously~\cite{Nayyar2011,Li2016,Fu2018}, quantitative differences exist depending on the nature of the hyperelastic model chosen.

\section{Far-from-threshold Analysis}

\subsection{The Cerda-Mahadevan model}
In 2003, Cerda \& Mahadevan (CM) proposed to study the wrinkle pattern in this system, and specifically the apparent reduction of the characteristic wavelength upon increasing the tensile load, by focusing on the system far from the instability threshold of the planar state. The rationale of this proposal was motivated by combining two observations: First, considering the planar state of a stretched Hookean sheet with Poisson ratio $\nu>0$, it is known that imposing tensile strain $\varepsilon$ induces a proportional amount of transverse compressive strain $\nu \varepsilon$. Second, planar sheets buckle when subjected to compressive strain that exceeds a threshold value $\varepsilon_c$, which vanishes with the sheet's thickness or more precisely, $\varepsilon_c(t_o/\Wo) \to 0$ as $t_o/\Wo \to 0$. Rather than focusing on the near-threshold regime ($\varepsilon \gtrsim \varepsilon_c$), which is the realm of classical pattern formation theory, CM proposed that observed wrinkle patterns in sufficiently thin sheets
may be understood by studying a Hookean far-from-threshold regime: 
\begin{equation} \varepsilon_c \ll \varepsilon \ll 1 \, . 
\label{eq:FT-regime}
\end{equation} 
In this regime, wrinkles are fully developed and affect the stress field in the sheet non-perturbatively, which means that one can no longer approximate it through the stress of the planar state which is unwrinkled. On the other hand, since for sufficiently thin sheets the threshold value 
$\varepsilon_c$ becomes arbitrarily small, far-from-threshold conditions may be realized even for $\varepsilon \ll 1$. This observation allows a quantitative study of this wrinkling phenomenon with linear (Hookean) elasticity and thereby FvK equations, for which the only source of nonlinearity is the geometric dependence of in-plane strain on out-of-plane deformations. 

The original CM model yielded two primary predictions, for the characteristic wavelength of wrinkles $\lambda$, and their amplitude $A$. According to their model, the wavelength is given by: 
\begin{equation}
\text{wavelength (CM):}  \ \ \ \lambda =  \frac{\sqrt{2 \pi L_o t_o}}{ [3(1 - \nu^2)\eps]^{1/4}}\,.
\label{eq:CM-wave}
\end{equation}
Importantly, this prediction for the wavelength $\lambda$ 
stems from analysis of periodic solutions to the FvK equations, and does not depend on the actual value of the amplitude $A$ of wrinkles (the only assumption made is that the wrinkled shape is of small slope, namely $A/\lambda \ll 1$). Regarding the amplitude, the model predicts: 
\begin{equation}
\text{amplitude (CM):}  \ \ \ A \sim \lambda \sqrt{\nu \varepsilon}  \ . 
\label{eq:CM-amplitude}
\end{equation}  
In order to understand the rationale underlying Eq.~(\ref{eq:CM-amplitude}), let us consider the wrinkle profile away from the clamped edges ({\emph{e.g.}} the center of the sheet) and denote by $\Wa$ the {\emph{actual}} arc-length of the sheet in the transverse ($\hat{y}$) direction, and by $\Wp$ the {\emph{projected}} width of the wrinkled sheet onto the plane. 
Assuming, $\Wa - \Wp \ll \Wp$, 
we have the purely geometric relation:
\begin{equation}
A \approx C \lambda \Phi \ \ , \ \ \Phi^2 \equiv {\frac{\Wa-\Wp}{\Wp}} \ ,  
\label{eq:amp-geo}
\end{equation} 
where $\Phi^2$ is the length (per wrinkle) that is ``wasted" by the undulations of the shape, and $C$ is a numerical constant that depends on the actual wrinkled shape. For instance, if the shape is a sinusoidal undulation $\zeta(y) = A \cos(2\pi y/\lambda)$, one obtains $C = 1/\pi$. The assumption underlying Eq.~(\ref{eq:CM-amplitude}) is that in the far-from-threshold regime: {\emph{(a)}} the compressive (transverse) strain vanishes, hence $\Wa = \Wo$, where $\Wo$ is the original width of the sheet in the undeformed state, and {\emph{(b)}} the planar projection of the stretched sheet continues to contract in width, even after the formation of wrinkles, according to the Hookean Poisson relation: $\Wp \approx \Wo (1-\nu\varepsilon)$.   


\subsection{The wavelength}

We tested the proposed scaling for the wavelength against our experimental data. The wavelength of the observed wrinkles was obtained by measuring the distance between neighboring central peaks in Fig.~\ref{fig:profiles}(a). In Fig.~\ref{fig:strain}(b), the wavelength $\lambda/\Wo$  normalized by the initial width of the sheet corresponding to the experiments and the MR model simulations are plotted as a function of applied tensile strain $\eps$. 
The data from our experiments and simulations are observed to collapse reasonably well with no fit parameters. Further, the overall decreasing trend with strain can be described by Eq.~\ref{eq:CM-wave} after multiplying by a factor 0.8 to the data and using $\nu = 0.5$ corresponding to a perfectly incompressible material. In examining the trends, it should be noted that the proposed scaling for the wavelength was derived assuming $\eps \gg \eps_c$. Consequently, discrepancy should be expected near threshold, and is indeed observed. Thus, we conclude that for the range of parameters addressed by our experiments and simulations the overall decrease in wavelength is captured reasonably well by the CM model. This conclusion is also similar to that reached by Nayyar, \textit{et al.}~\cite{Nayyar2011} based on simulations with NH model. 

Although the data in Fig.~\ref{fig:strain}(b) provides support to the prediction~Eq.\,(\ref{eq:CM-wave}), even for values of the applied strain that are beyond the asymptotic range~Eq.\,(\ref{eq:FT-regime}) assumed by the CM model, we note that in Fig.~\ref{fig:strain}(b) the normalized wavelength $\lambda_o/\Wo$ is in the interval $(0.9,0.17)$, and it is not obvious that this data is sufficient to confirm the validity of the scaling law~Eq.(\ref{eq:CM-wave}), ({\emph{i.e.}} in the relevant asymptotic limit $\varepsilon \ll 1$. More generally, Eq.~(\ref{eq:CM-wave}) is a particular example of a broader scaling law, $\lambda \sim (B/K_{\rm eff})^{1/4}$, 
which was proposed already in Ref.~\cite{Cerda2003} and expanded in Ref.~\cite{Paulsen2016}, whereby the wrinkle wavelength derives from a balance between the bending modulus $B$ of the solid sheet and an effective stiffness $K_{\rm eff}$. The effective stiffness may originate from a real substrate attached to the sheet ({\emph{e.g.}} a liquid bath or a compliant solid foundation), a global curvature that may be imposed on the sheet, or a tension $T$ exerted along wrinkles. In the latter case, the effective stiffness is $T/\ell^2$, where $\ell$ is a characteristic scale in the non-oscillatory direction over which the wrinkle amplitude varies. In our case, there is no actual substrate nor global curvature (since the wrinkled sheet remains close to the original plane), and Eq.~(\ref{eq:CM-wave}) follows immediately if one assumes that the characteristic longitudinal scale $\ell \sim L_o$. We note that the data plotted in Fig.~\ref{fig:strain}(b) may be actually consistent with $\ell \sim L_o^{\alpha} \Wo^{1-\alpha}$, with $\alpha \neq 0$. In order to test the dependence of the longitudinal scale $\ell$ on the aspect ratio $L_o/\Wo$ one would have to perform experiments and/or simulations on a broad range of $L_o/\Wo$, which is beyond the scope of the current work.       


\begin{figure*}
\begin{center}
\includegraphics[width=13cm]{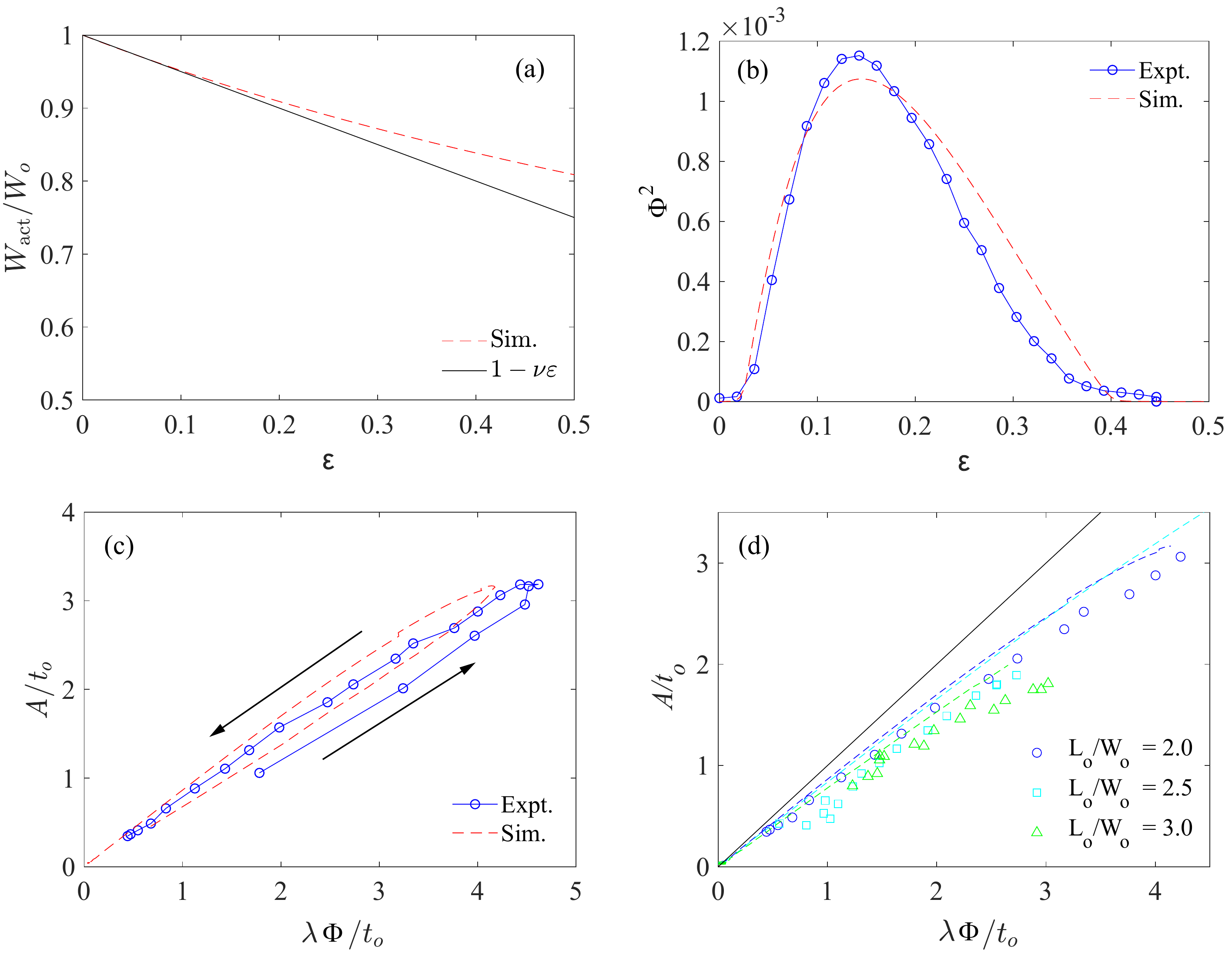}
\end{center}
\caption{(a) The actual width $\Wa$ (at the center of the sheet, $x=0$), extracted from simulations, plotted as a function of the applied tensile load, $\varepsilon$ ($L_o/\Wo = 2$). It can be observed that $\Wa$ 
decreases consistent with $\Wa = \Wo (1-\nu \varepsilon)$, indicating that the collapse of compressive stress (rather than compressive strain) characterizes both the planar state (Poisson effect) as well as the fully developed, far-from-threshold, wrinkled state of the sheet. Nonlinear elastic effects -- similar in nature to those observed in the stress-strain response shown in Fig.~\ref{fig:props} -- lead to deviations at $\varepsilon > 0$.  
(b) The average difference $\Phi^2$ between the actual arc-length $\Wa$ and its planar projection $\Wp$ (plotted at the center of the sheet, $x=0$) 
increases and then decreases as the wrinkles 
exhibits a non-monotonic dependence on the applied tensile load $\varepsilon$, as wrinkles appear and then disappear with increasing $\varepsilon$. (c)  The normalized amplitude $A/t_o$ {\emph{versus}} $\Phi \lambda/t_o$
for $L/W =2$ plotted over the whole range of the applied tensile load $\varepsilon$, when the measured strain is greater than experimental errors. The arrows indicate the trajectory of the points as $\varepsilon$ is increased. (d) Amplitude for large strain $\eps > 0$  and various $L_o/\Wo$ for $\eps > 0.15$.} 
\label{fig:wrinkle}
\end{figure*}

\subsection{The amplitude}
In contrast to the prediction of the CM model for the wavelength, Eq.~(\ref{eq:CM-wave}), whose consistency with experimental data and simulations has been noted already \cite{Nayyar2011}, the model prediction for the wrinkle amplitude, Eq.~(\ref{eq:CM-amplitude}), appears to be inconsistent with observations. Notably, Eqs.~(\ref{eq:CM-wave},\ref{eq:CM-amplitude}) predict that for a given sheet ({\emph{i.e.}} fixed values of $t_o,\Wo,L_o$) the amplitude scales as $\sim \varepsilon^{1/4}$ upon increasing the tensile load. However, experiments and simulations exhibit the opposite trend, whereby the amplitude starts to decrease already at rather small values of $\varepsilon$.  
Furthermore, the very reasoning underlying the transition from the geometric relation, Eq.~(\ref{eq:amp-geo}), to Eq.~(\ref{eq:CM-amplitude}), is unclear. While the planar state exhibits transverse compressive stress ($\sigma_{yy} <0$) at some zones close to the clamped edges, a compressive transverse strain $\varepsilon_{yy} \approx -\nu \varepsilon$ is necessary to eliminate any compressive stress in the bulk, and thereby the elastic energy that would stem from such a nonzero stress component. In fact, this is the precise logic underlying ``Poisson effect." An ``inextensibility" condition assumed in the CM model to characterize the far-from-threshold behavior ($\varepsilon_{yy}=0 \Rightarrow \Wa=\Wo$) would therefore give rise to finite transverse compression in the bulk of the sheet ($\sigma_{yy} \approx -\nu Y \varepsilon$), which would entail an elastic energy that is even higher than the energy of the planar state. The experimental and numerical observations, along with the conceptual difficulty in Eq.~(\ref{eq:CM-amplitude}),   
led workers in this field \cite{Nayyar2011,Healey2013} to attribute the wrinkling instability in this system to effects of non-Hookean elasticity.    

A deeper perspective on the wrinkle amplitude requires us to address first the basic principles of the mechanics in the far-from-threshold regime Eq.\,(\ref{eq:FT-regime}). While the nontrivial effect of clamping the short edges is crucial for understanding the origin of compression and wrinkling in our problem, there exist other set-ups that exhibit a similar type of tension-induced wrinkling, but are amenable to analytic solution due to their inherent axial symmetry. A notable example is the Lam\'e set-up of an annular sheet that exhibits radial wrinkles when subjected to distinct tensile loads at its inner and outer radii~\cite{Davidovitch2011}. Studies of the far-from-threshold mechanics in this and other systems \cite{Chopin2015,Vella2018b,Davidovitch2019} revealed three important principles: 

{\emph{(a)}} It is the transverse stress ({\emph{i.e.}} stress component in the direction of wrinkly undulations) rather than strain, whose ``collapse" is enabled by the formation of wrinkles.    

{\emph{(b)}} The collapse of compressive stress requires a specific displacement of material lines, which is obtained by solving the corresponding tension field theory equations~\cite{Stein1961,Pipkin1986,Steigman1990}. In problems of axial symmetry~\cite{Davidovitch2011,Vella2018b,Davidovitch2019}, this implies a certain dependence of the radial displacement ($u_r(r)$) on the exerted tensile load imposed on the sheet ({\emph{e.g.}} Eqs.~13,17 in Ref.~\cite{Davidovitch2011}).  

{\emph{(c)}} Conditions {\emph{(a,b)}} underlie a ``slaving condition", which connects the amplitude-wavelength ratio to the excess length 
that must be ``wasted" by the wrinkling undulations, to enable the desired collapse of the compressive stress. In axisymmetric set-ups, this excess length was called a ``confinement function," $\Phi^2(r) = u_r(r)/r$  
 ({\emph{e.g.}} Eqs.~14,15 in Ref.~\cite{Davidovitch2011} and Eq.~2 in~\cite{Paulsen2016}).

Although the complexity of our problem does not allow for analytic solution of tension field theory, we recognize Eq.~(\ref{eq:amp-geo}) as the amplitude-wavelength slaving condition, whose satisfaction guarantees the collapse of compressive stress in the wrinkled zone -- item {\emph{(c)}} in the above list. The collapse of compressive stress, corresponding to item {\emph{(a)}} in the above list, implies that the actual arc-length, $\Wa$, retains its value in the analogous planar state ($\Wa = (1-\nu \varepsilon)$, which is needed to eliminate transverse compression in the presence of tensile strain $\varepsilon$. Finally, a ``confinement function", $\Phi^2(x) = (\Wa - \Wp)/\Wp $, 
which is the analog of $\Phi^2(r) = u_r/r$ for problems of axial symmetry (and whose $x$-dependence is omitted in Eq.~(\ref{eq:amp-geo}) for simplicity), derives from a suitable solution of the equations of tension field theory -- item {\emph{(b)}} in the above list. The differences between this implementation of far-from-threshold theory and the assumptions made in the original CM model \cite{Cerda2003} are summarized in Table 1.     

\begin{table}[ht]
  \begin{center}
    \label{tab:table1}
    \begin{tabular}{|c|c|c|} 
    \hline
    \hline
      \textbf{} & \textbf{Current version} & \textbf{Original CM model}\\
      \hline 
      $\Wa$ & $\Wo (1-\nu \varepsilon)$ &  $\Wo$ \\ \hline 
      $\Wp$ & tension field theory  & $\Wo (1-\nu \varepsilon)$ \\ \hline
    \end{tabular}
  \end{center}
    \caption{Implementation of the ``slaving condition," Eq.~(\ref{eq:amp-geo}).}
\end{table}

In order to test our proposed revision of the CM model, namely Eq.~(\ref{eq:amp-geo}) with $\Wa,\Wp$ in Table I, we first extract from our simulations the actual width $\Wa$ of the wrinkled sheet at increasing values of the applied tensile load $\varepsilon$. The plot, Fig.~\ref{fig:wrinkle}(a), for $L_o/\Wo = 2$, shows that the decrease of $\Wa$ is given by $\Wa = \Wo (1-\nu \varepsilon)$ as wrinkles develop for $\varepsilon <0.1$. At greater $\varepsilon$, nonlinear effects become important leading to deviations from the linear decrease, complementary to the nonlinear stress-strain response in Fig.~\ref{fig:props}. This decrease implies the absence of transverse compressive stress ($\sigma_{yy} = 0$) and the consequent presence of transverse compressive strain, even beyond the Hookean regime of small strains. 

Next, we plot in Fig.~\ref{fig:wrinkle}(b) the evolution of $\Phi^2$ as a function of the applied tensile load $\varepsilon$ measured across the central width of the sheet, i.e. $x=0$.  Below the threshold $\varepsilon_c$ for wrinkling instability, the projected and actual width are trivially equal, such that $\Phi^2 = 0$. For $\varepsilon > \varepsilon_c$, we observe that $\Phi^2$  
increases and then decreases, eventually becoming zero beyond a finite value of $\varepsilon$ (above which wrinkles vanish and the planar state become stable again). The important lesson to draw from this figure is that the actual behavior of $\Phi^2$, which derives from a solution of tension field theory, is dramatically different, being an order of magnitude smaller, from the one assumed by the original CM model ($\Phi^2 \approx \nu \varepsilon$) \cite{Cerda2003}. 
Specifically, the confinement $\Phi^2$ exhibits a non-monotonic dependence on $\varepsilon$ even for rather small values of $\varepsilon$, at which the mechanics is still well approximated by Hookean mechanics (see Fig.~\ref{fig:strain}.)     

In Fig.~\ref{fig:wrinkle}(c), we plot the measured amplitude $A$ {\emph{versus}} the measured values of $\lambda \Phi$,  for $L_o/\Wo =2$. It should be noted that $A/t_o$ increases first as $\varepsilon$ is increased, and then decreases after reaching a maximum value with the large strain data appears in the bottom left corner, close to the origin as indicated by the arrows. We find that the experimental data and the MR model simulations match well, and both agree with the geometric relation, Eq.~(\ref{eq:amp-geo}) with a single numerical constant $C =1$, even at large strains. 

In Fig.~\ref{fig:wrinkle}(d), we show the same analysis with three different aspect ratios in the experiments (markers) and corresponding MR model simulations (dashed curves) that span the range over which wrinkles are observed. The collapse of the data is remarkable considering the relatively small amplitudes of the wrinkles which are here of order of the thickness of the sheet. The somewhat systematic shift of the experimental data points to the right relative to the numerical curves can be attributed to the fact that there is higher cumulative error which occurs in obtaining $\Phi$ from the experimental data compared with $A$. The theoretical prediction 
of Eq.~(\ref{eq:amp-geo}) with best-fit fixed value of the constant 
$C =1$ corresponding to the highest applied strains shown by the solid line. The systematic deviations observed between the data points and solid line indicates that the wrinkle profile slightly varies upon increasing the applied tensile load, affecting a weak $\varepsilon$-dependence (namely $C \rightarrow C(\varepsilon)$) in Eq.~(\ref{eq:amp-geo}). 

\section{Conclusions} 

We have demonstrated with experiments that a sufficiently thin elastic sheet wrinkles when stretched over a finite range of strains. We provided clear evidence that the wrinkle amplitude first grows upon increasing applied strain, but then decreases as the strain is continued to be increased. By using simulations of the MR model, we showed that all the trends observed in the experiments are observed with this model within experimental error, once the material parameters are appropriately calibrated. The wrinkles are observed at higher strains in comparison with SVK and NH models.

While non-Hookean mechanics is needed to explain certain aspects of the wrinkle pattern, most notably the persistence of wrinkles to applied strain $\varepsilon \approx 0.4$, the basic aspects of the phenomenon are describable by far-from-threshold analysis, which employs the asymptotic vanishing of the instability threshold with the sheet's thickness ($\varepsilon_c \to 0$), and addresses the asymptotic regime Eq.\,(\ref{eq:FT-regime}) through Hookean elasticity. This approach was pioneered by the original CM model \cite{Cerda2003}, which yields a powerful prediction for the wavelength $\lambda$, but the ``inextensibility" assumption ({\emph{i.e.}} vanishing transverse compressive strain, $\varepsilon_{yy} =0$) that underlies the model prediction of wrinkle amplitude that increases with tensile load is revoked by an opposite trend observed in experiments and simulations. We showed that this assumption must be replaced by the collapse of transverse compressive stress ($\sigma_{yy} =0 \Rightarrow \varepsilon_{yy} =-\nu\varepsilon$), such that the amplitude-to-wavelength ratio is determined by a ``confinement function", $\Phi^2(x)$, whose dependence on the applied tensile load $\varepsilon$ is determined by a corresponding solution of tension field theory. In set-ups of high degree of symmetry,  
analytic solutions of tension field theory have been found and were shown to provide a basis for a far-from-threshold analysis of wrinkle patterns   \cite{Davidovitch2011,Chopin2015,Vella2018b,Davidovitch2019}. In the current set-up, despite the deceptive simplicity of the wrinkle pattern, the boundary conditions implied by combining clamping (at the short edges) and free boundaries (the long edges), require one to obtain a numerical solution of the tension field theory; our experiments and simulations show that such a solution results in a nontrivial confinement function $\Phi^2(x)$, whose evaluation at $x=0$ is shown in Fig.~\ref{fig:wrinkle}(b). We hope that our findings will motivate a systematic effort to develop far-from-threshold analysis in non-symmetric set-ups and beyond the Hookean asymptotic regime given by Eq.\,(\ref{eq:FT-regime}).                          

\vspace{2cm}

\begin{acknowledgments}
We thank Jose Bico for illuminating conversations. This work was supported by the National Science Foundation under grants DMR 1508186 (A.P. and A.K.) and DMR 1822439 (M.X. and B.D.). Simulations were performed by using the UMass cluster  at the Massachusetts Green High Performance Computing Center. 
\end{acknowledgments}

\bibliography{elastic}

\end{document}